\documentclass[aps,pre,onecolumn,superscriptaddress]{revtex4}
\usepackage[dvips]{graphicx}
\usepackage[dvips]{color}
\usepackage{breakurl,amsmath,amssymb,pst-node,eepic}

\usepackage{subfigure}
\usepackage{tikz}
\usepackage{amsmath}
\usepackage{amsfonts}
\usepackage{amssymb}
\usepackage{multirow}
\usetikzlibrary{shapes, arrows, calc, positioning,matrix}
\usepackage{bbold}

\begin{document}

\title{Infectivity Enhances Prediction of Viral Cascades in Twitter}

\author{Weihua Li}
\email{weihuali89@gmail.com}
\affiliation{LMIB, BDBC and School of Mathematics and Systems Science, Beihang University, Beijing 100191, China}
\affiliation{Department of Computer Science, University College London, 66-72 Gower Street, London WC1E 6EA, UK}
\affiliation{Systemic Risk Centre, London School of Economics and Political Sciences, Houghton Street, London WC2A 2AE, UK}

\author{Skyler J. Cranmer}
\affiliation{Department of Political Science, The Ohio State University, Columbus, OH 43210, USA}

\author{Zhiming Zheng}
\affiliation{LMIB, BDBC and School of Mathematics and Systems Science, Beihang University, Beijing 100191, China}

\author{Peter J. Mucha}
\affiliation{Department of Mathematics, The University of North Carolina, Chapel Hill, NC 27599, USA}


\begin{abstract}
Models of contagion dynamics, originally developed for infectious diseases, have proven relevant to the study of information, news, and political opinions in online social systems.
Modelling diffusion processes and predicting viral information cascades are important problems in network science.
Yet, many studies of information cascades neglect the variation in infectivity across different pieces of information.
Here, we employ early-time observations of online cascades to estimate the infectivity of distinct pieces of information.
Using simulations and data from real-world Twitter retweets, we demonstrate that these estimated infectivities can be used to improve predictions about the virality of an information cascade.
Developing our simulations to mimic the real-world data, we consider the effect of the limited effective time for transmission of a cascade and demonstrate that a simple model for slow but non-negligible decay of the infectivity captures the essential properties of retweet distributions.
These results demonstrate the interplay between the intrinsic infectivity of a tweet and the complex network environment within which it diffuses, strongly influencing the likelihood of becoming a viral cascade.
\end{abstract}

\maketitle

Massive data sets that comprehensively capture users' behaviours in online social systems and their underlying network structures have reached an unprecedented scale, making it possible to develop computational methods to model complex patterns of human behaviour at both individual and population levels \cite{castellano2009statistical, monsted2017evidence, muchnik2013social}.
Among various human-induced online processes, the study of social contagion---the spread of information, ideas, and behaviours through social networks---has attracted tremendous attention, especially in the fields of computational social science and network science \cite{ugander2012structural, lazer2009life}.
Many studies examine these peer-to-peer diffusion processes by focusing on a single piece of information and making assumptions about infectivity, recovery probabilities, and their intrinsic relations to network structures \cite{goffman1964generalization, daley1964epidemics, pastor2001epidemic, barrat2008dynamical, vosoughi2018spread}.
We consider measuring the infectivity of information cascades to be the crux for predicting their ultimate virality.

\begin{figure}[!htpb]
	\centering
	\includegraphics[width=0.98\linewidth]{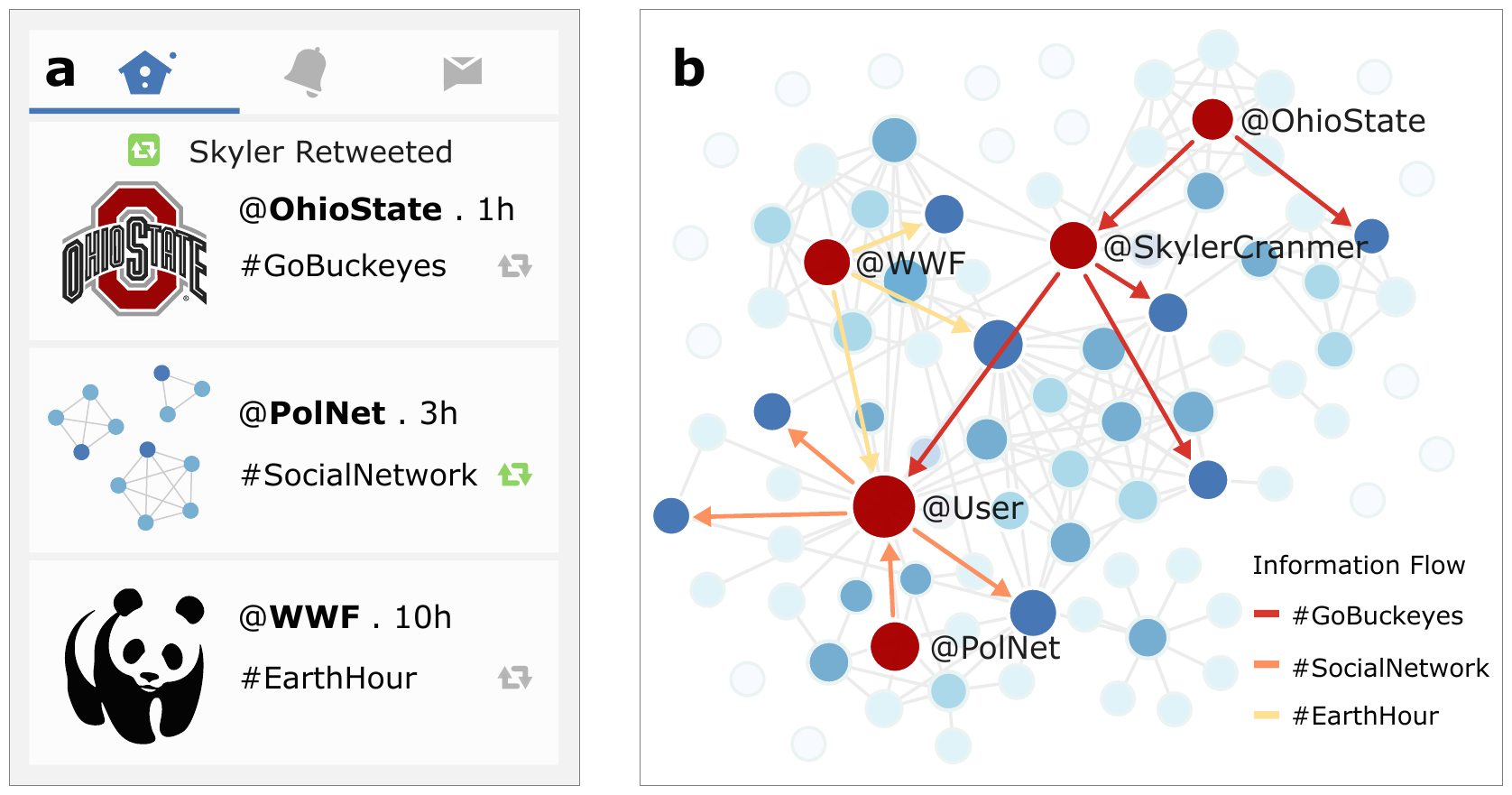}
	\caption{\textbf{Schematic of social contagion information diffusion in Twitter.}
		\textbf{(a)} The Twitter user interface that displays three latest tweets with different degrees of interestingness from her friends.
		The first message was originally posted by someone with whom she does not have direct connection, but she is still able to see it after being retweeted by one of her friends.
		She chose to retweet the second tweet she found interesting, extending the information flow of that message to all her followers.
		If the ``memory length" of this user is $3$, she will not read or retweet messages posted more that 10 hours ago (the time of the third item in the display).
		\textbf{(b)}, The online network environment of involved users and the flows of information cascades.
	}
	\label{Twitter}
\end{figure}

Previous research has successfully advanced the modelling of information spread by studying memes in Twitter data, where a meme is defined by the use of a hashtag and includes all of the tweets with that hashtag \cite{weng2012competition,weng2013virality,gleeson2014competition,weng2014predicting,qiu2017limited}.
Here, we reanalyze these data with an exclusive focus on modelling the direct transmission of information through a social network in the form of retweets.
Our reason for focusing on retweets is that the transmission of a particular hashtag is more likely to occur not only from person to person through online social ties \cite{weng2014predicting}, but also through a broadcasting manner across other media outside the specific social network.
As observed in Ref.~\citenum{goel2015structural}, broadcasts contribute substantively to viral events, e.g., the World Cup Final attracts about $1$ billion viewers worldwide, while news coverage from popular websites also reaches a similar number of Internet users.
In such popular events, the discussion of a meme in broadcasting media (e.g.~social network platforms, TV shows, radio and news reports) can greatly boost its spread.
Retweets, by contrast, constitute an information cascade that originated from an identifiable individual user and is a contagion spread mostly through the links of the follower network (Fig.~\ref{Twitter}).

The Twitter data we use contains a follower network with $6.0 \times 10^5$ users, $1.7 \times 10^6$ retweets and $1.2 \times 10^7$ tweets generated by these users in $33$ days \cite{weng2013virality, weng2014predicting}.
We estimate the probability distribution of the infectivity of cascades from real data, and simulate the process on the follower network (see Methods).
A cascade consists of retweets that have the same hashtag and the same user who initially posted the tweet, together with the tweet that originated the cascade.

Previous studies have demonstrated that the topology of networks, especially the community structure, has pronounced effects on information diffusion \cite{rosvall2008maps,weng2013virality}.
Communities could promote spread by homophily and social reinforcement, but may also hinder wider spread by trapping information, resulting in a high concentration of retweets within a community.
To examine the influence of community structures, Weng et al.\ \cite{weng2013virality} introduced two statistical features of memes, which we modify for retweet cascades: the adoption dominance $g$ computes the proportion of users retweeting the cascade in the community with the most adopters; and the retweet entropy $H^{r}$ quantifies the distribution of retweets across different communities, as a measure of the concentration of the cascade across communities.
We compute both measures based only on retweets in their early stages (first $50$ tweets) to avoid bias from a cascade's popularity.

\begin{figure}[!htpb]
	\centering
	\includegraphics[width=0.66\linewidth]{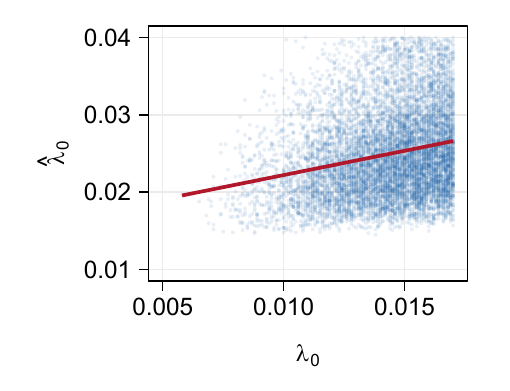}
	\caption{\textbf{Comparison between real and estimated infectivity in simulations.} Real infectivity $\lambda_{0}$ and estimated infectivity $\hat{\lambda}_{0}$ computed from simulation data according to equation \ref{mt} without considering the decay effects. The solid line is the linear regression fit. Estimates are calculated from only the first $50$ retweets of each tweet, so that they may be used to try to predict whether a given cascade ``goes viral."}
	\label{fig:infectivity}
\end{figure}

Retweet cascades are very different from hashtag memes in that we can more realistically assume that social contagion through the follower network is the major mechanism by which the retweet cascade is propagating.
To provide direct evidence of this, we sampled $10 ^ 5$ tweets and retweets, respectively, finding that for $23.8\%$ of tweets we can find at least one earlier tweet with the same hashtag from the user's friends, while $46.0\%$ of retweets have at least one friend who previously retweeted in the same cascade.
Importantly, these percentages are limited by the specific follower network available in the data set, which inherently undercounts the possibility of transmission through the online social network because the network in the data only includes the reciprocal following ties (to better reflect real social relations).
We estimate the infectivity of a specific cascade assuming that all such identifications are the actual paths of information transmission, using only the first $50$ retweets (see Methods).
Despite the relatively high inaccuracies observed between the true and predicted infectivities in our simulated data (where we know the true imposed infectivity, cf. real Twitter data), we note the overall trends of the infectivity estimates are in the right direction, with a slope of $0.92$ and $R^2 = 0.05$ (Fig.~ \ref{fig:infectivity}).
We thus proceed to consider predictive models for virality that include such estimates of cascade infectivity.

We now test whether this simple model of infectivity demonstrates predictive power for identifying viral retweet cascades in real Twitter data.
In Ref.~\citenum{weng2013virality}, Weng et al.\ used community concentration features to predict viral memes with three models: the random guess (RG) model randomly samples the cascade without any predictors;
the null model (NM, referred to as the ``community-blind model" in Ref.~\citenum{weng2013virality}) employs the number of distinct users and the total number of neighbours of early retweet users;
the community-based (CB) model also incorporates three community-based features of the Twitter network: the number of infected communities, retweet entropy $H^r$, and the fraction of intra-community user interactions (see Appendix \ref{si:6}).
We introduce two additional models adding features to the NM model to predict viral cascades with infectivity estimates: the infectivity-based (IB) model uses the estimated rate of infectivity $\hat{\lambda_0}$ from equation (\ref{mt}),
where $\langle k \rangle$ is the mean degree of early retweet users;
and the community \& infectivity based (C\&I) model combines all of these infectivity and community-based features.
Each of our classifiers includes only information about the first $50$ retweets of each tweet, to try to predict whether the retweet cascade ``goes viral". We train random forest classifiers on $1,272$ real Twitter cascades and $20,000$ simulated cascades sampled from $20$ replications, using $10$-fold cross validation to predict viral cascades that attract more retweets than a certain percentile threshold $\theta$ of all cascades.

\begin{figure}[!htpb]
	\centering
	\includegraphics[width=0.96\linewidth]{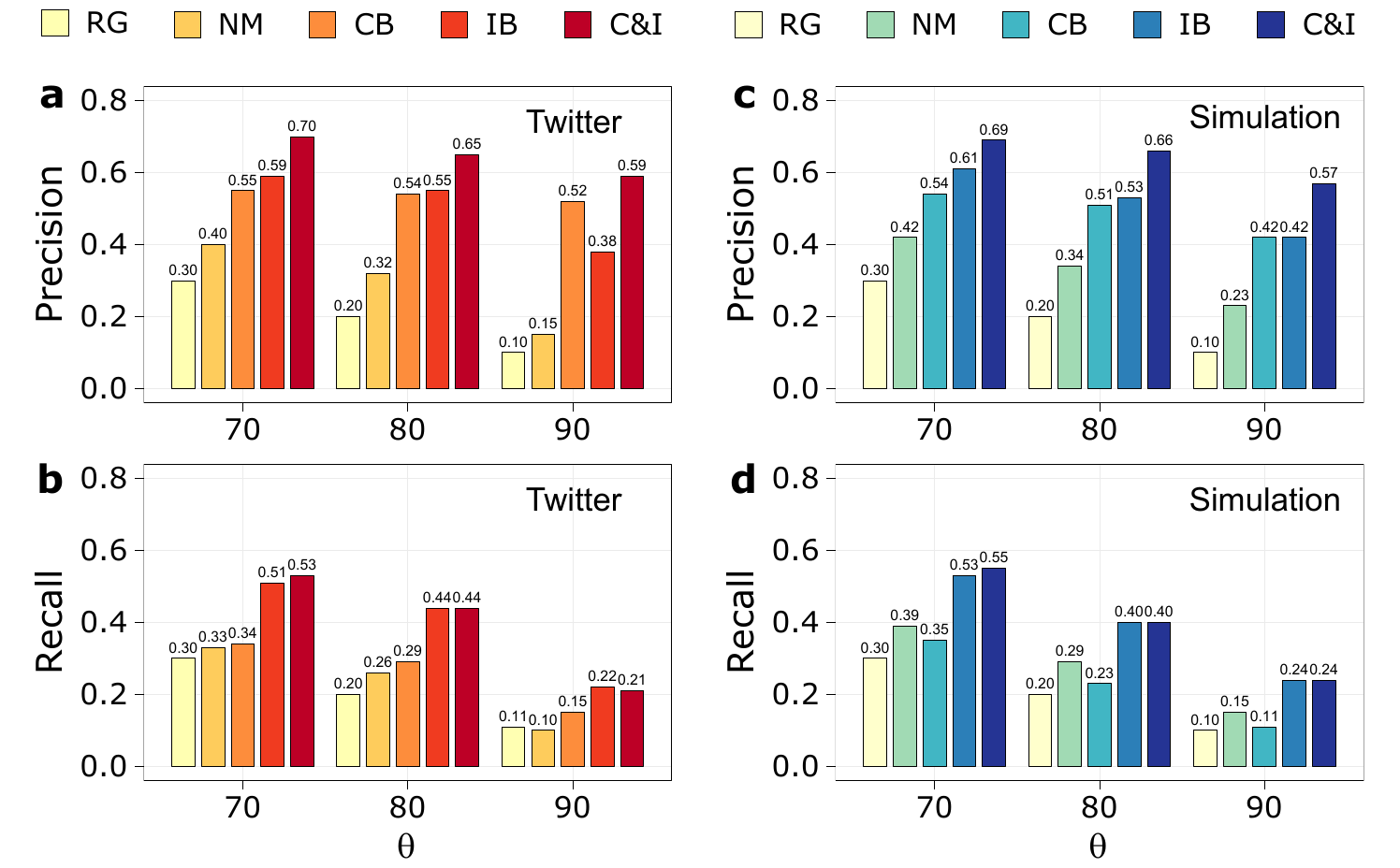}
	\caption{\textbf{Random forest model predictions.} We predict whether a cascade will go viral or not; a cascade is viral if it produces more retweets than a certain percentile threshold ($\theta = 70, 80, 90$) of cascades, using community-based features and infectivity estimates that are calculated based on the initial $50$ retweets for each cascade. Random forests are trained on sets of features delineated by the labels RG, NM, CB, IB and C\&I (see the main text). The classifier including estimated infectivity (IB) typically outperforms the community-based model (CB), while combining all of the community-based and infectivity features (C\&I) gives the best predictions overall. \textbf{a}, Precision rates of Twitter data. \textbf{b},  Recall rates of Twitter data. \textbf{c}, Precision rates of retweet data from simulations. \textbf{d}, Recall rates of retweet data from simulations.}
	\label{plot_rf}
\end{figure}

The results on the Twitter data suggest that in most cases our IB model performs better than the CB model (Fig.~\ref{plot_rf}ab), indicating that estimated infectivity alone can improve the prediction even more than the community-based predictors.
Moreover, the C\&I model, incorporating both community and infectivity factors, reveals a striking increase of predictive power above the other models.
Fig.~\ref{plot_rf}cd shows random forest model prediction and recall rates on retweet data generated by our simulations, indicating patterns consistent with those observed in the Twitter data.
The IB model, only adding infectivity to the NM model, is comparable to the CB model that includes three community features, and by considering all predictors the C\&I model excels in both precision and recall rates.
We note that replacing the estimated $\hat{\lambda}_0$ by the true $\lambda_0$ used in the simulations---a test we can obviously not reproduce in the real Twitter data---yields additional improvement in classification (Table \ref{table:rf}), suggesting substantial potential for a more refined estimate of $\hat{\lambda_0}$ to lead to even greater accuracy for predicting viral cascades.

We further test our results using logistic regression with the same set of features as in the C\&I model.
We find that estimated infectivity is still a significant predictor in simulation data, but not in predicting virality in the real Twitter data (Tables \ref{table:logit,table:logitSimLambar, table:logitSimLam}).
There may be multiple reasons for this apparent discrepancy between the random forest and logistic regression results.
One possibility is that logistic regression is too specific in the functional form in which it estimates the probability of virality. In particular, we note the substantial noise in estimating infectivity we observe in our simulations; without any way to compare the estimated infectivities with ``true" values in the real Twitter data, we cannot know whether the effect of this noise interacts poorly with the log-odds-shift assumptions of logistic regression.

\begin{figure}[!htpb]
	\centering
	\vspace{-1.2cm}
	\includegraphics[width=0.96\linewidth]{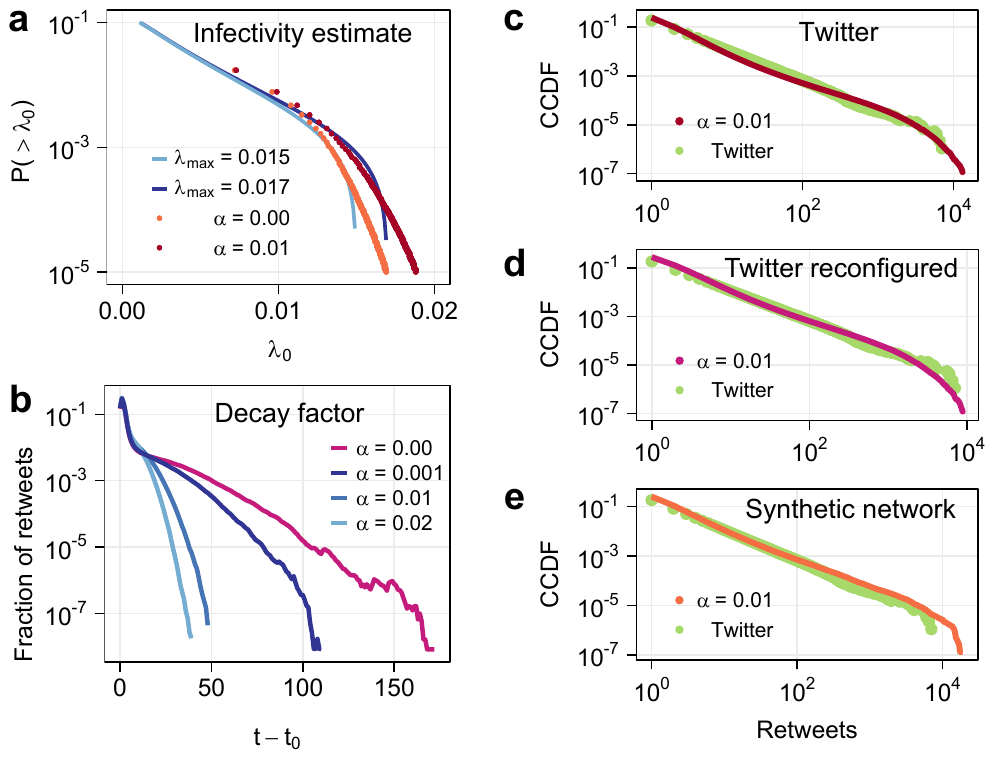}
	\caption{\textbf{Simulation parameter settings and results.}
		\textbf{a}, Truncated lognormal fit. Light and dark blue lines are fit with theoretical distribution function equation (\ref{p_theory}), and the red and orange points are fit with distribution computed from real Twitter data with equation (\ref{A_t}). The parameters used in equation (\ref{p_theory}) are as follows: when decay factor $\alpha = 0$, $\mu = \ln 0.0012, \sigma = \ln 2.4, \lambda_{\mathrm{max}} = 0.015$; when $\alpha = 0.01$, $\mu = \ln 0.0012, \sigma = \ln 2.4, \lambda_{\mathrm{max}} = 0.017$.
		\textbf{b}, Retweets at time $t$ of cascades originated at time $t_0$ with different decay factors.
		\textbf{c-e}, Complementary cumulative distribution functions (CCDFs)---the fraction of cascades with more than $n$ retweets for numerical simulations, compared with retweets from empirical Twitter data marked by green points. The model parameters are identical except for the network structure: \textbf{c}, The empirical Twitter follower network with $N = 5.95 \times 10^5$ and $\langle k \rangle = 47.94$; \textbf{d}, Reconfiguration of the empirical Twitter network preserving the degree distribution; \textbf{e}, Scale-free network with $N = 5 \times 10^5$, $\langle k \rangle = 48$ and exponent $\gamma = 2.8$. }
	\label{main_results}
\end{figure}

Our simulations emulate the real-world diffusion process in Twitter by taking into consideration several human behavioural factors, such as a limited memory length and a gradual decrease in interest, in a simplified simulation model.
We estimate a fixed memory length for all users from data and additionally incorporate a small but non-zero decay parameter to the infectivity of each cascade (see Methods).
The initial infectivities of cascades are sampled from a probability distribution computed from empirical data (Fig.~\ref{main_results}a).
The decay effect mainly affects the long time dynamics of viral cascades (Fig.~\ref{main_results}b).
If we ignore the decay effect of infectivity, cascades with large infectivity will still keep spreading after long periods of time, even with fixed user memory length.
With a small but non-zero decay parameter $\alpha$, even the most popular cascades will diminish at some point, and the system quickly reaches equilibrium.
We then use simulations on networks with different structural properties but otherwise identical parameter settings to calculate the distributions of cascade sizes.

Fig.~\ref{main_results}c shows that our simulations on the Twitter follower network replicate well the cascade distribution observed in the data.
{We also run a simulation on a configuration model network with the same degree distribution as the empirical Twitter network (Fig.~\ref{main_results}d).}
Simulation results on a synthetic network generated by the algorithm in Ref.~\citenum{goh2001universal} with the power-law exponent $\gamma = 2.8$, representing an analogous degree heterogeneity of the Twitter network (see Figure \ref{deg_distribution}), also recover the statistical features of Twitter data (Fig.~\ref{main_results}e).
When we switch the decay parameter to $0.001$ and $0.02$, respectively, we still replicate the empirical retweet distribution fairly well by changing the corresponding $\lambda_{\mathrm{max}}$ parameter (Appendix \ref{si:5}).

We have demonstrated the predictive power of infectivity for identifying viral retweet cascades in real-world Twitter data and in simulation.
An important assumption of this study is that the spread of retweet cascades resembles the peer-to-peer social contagion through the Twitter follower network, which we argue is different from viral memes represented by hashtags that more heavily rely on transmission through broadcasting.
We demonstrate that the early spread rate for retweet cascades can be a good indicator of the intrinsic interestingness of a tweet, and that the corresponding estimate of infectivity gives improved prediction of virality.
But, importantly, the same scheme might not readily apply to some memes that need to be broadly broadcast before they become viral.
This difference may help explain why the measure of early infectivity of a hashtag in Ref.~\citenum{weng2014predicting} does little to improve the prediction of viral memes.

Our mean-field method to estimate infectivity from empirical data clearly leaves plenty of room for improvement.
The predictive ability of machine learning methods improves further on simulation data when we include the true infectivity, demonstrating the importance of accurate estimations of the cascade infectivity.
Apart from this indirect approach with strong assumptions, we could also design a more straightforward method.
The biggest challenge for such a measurement is to gather large-scale, high-quality data with which it is possible to infer accurate retweet relations.
Better data and more reliable methodology to estimate infectivity are key to improving the predictive power.

Our study shows that infectivity improves the prediction of viral cascades that are mostly induced by contagion along the links representing social network connections.
Network community structure captures additional local environmental factors such as homophily, social reinforcement and a trapping effect that further affect the spread and likelihood of virality of retweet cascades.
Nevertheless, the infectivity determines the internal attractiveness and seems to be one of the most important factors in driving the virality of a cascade.
Said another way, we have successfully demonstrated that the inherent quality of content---in the sense of being sufficiently interesting to have high infectivity---is an essential element promoting the chances of a successful spread that might not otherwise be as plausible in light of the local environmental factors.

\appendix

\section*{Methods}
\subsection{Data.}
The Twitter data, studied previously in Refs.~\citenum{weng2013virality} and \citenum{weng2014predicting}, comprise a reciprocal follower network of $N = 595,460$ nodes and the time-stamp record of $N_{twt} = 12,054,205$ tweets, of which $N_{ret} = 1,687,704$ are retweets, within a total time frame of $T = 33$ days and we treat a day as the time step. 
The data were collected in three data sets: (1), a reciprocal follower network where each edge is a pair of Twitter users who are following each other; (2), tweet timeline data with the hashtags and their adopters sorted by timestamp; (3), the retweet timeline data where each line is a hashtag followed by the sequence of its adopters retweeting about this hashtag from other users sorted by timestamp.
Note that the retweet data set is a subset of the tweet data set.

\subsection{Generating functions.}
The modeling of human factors---specifically a dynamical process with limited user memory length---can help to unveil the core features of contagion in complex social systems driven by peer-to-peer influence.
At every time step, a user generates a new tweet with innovation probability $\beta = (N_{twt} - N_{ret}) / N T$.
The infectivity $\lambda_0$ of a cascade is the probability that a follower will retweet it in one time step.
Let us consider the dynamical process of retweeting in more detail by focusing on a given information cascade with infectivity $\lambda_0$, posted online at time $t=0$, assuming for simplicity that all other cascades have infectivity equal to its mean, $\langle \lambda_0 \rangle$. We denote the distribution of retweets at time $t$ by $q_n(t)$, which is the probability that a cascade has popularity $n$ at $t$.
Following the probability generating function (PGF) formalism in Refs.~\citenum{wilf2013generatingfunctionology, newman2001random, gleeson2014competition}, we define the cascade PGF, parameterized by $x$, to be $H(t, x) \equiv \sum_{n=1}^\infty q_n(t)x^n$.
We assume the in-degree of all nodes to be $\langle k \rangle$, and characterize the heterogeneity of the out-degree distribution with PGF $f(x) \equiv \sum_{k=0}^\infty p_k x^k$, where $p_k$ is the probability of a node with out-degree $k$.
We seek to quantify $G(t, x)$ as the PGF for the retweet distribution at time $t$ of a random cascade branch that originates from a single user randomly chosen from a given cascade.
For the user and all of her followers, a tweet event increases the popularity of the given cascade by $1$, and places it at the top of the memory length window.
As a result, the PGF for the number of tweets at time $t$ is given by \cite{gleeson2014competition} $H(t,x) = x G(t,x) f(G(t,x))$.
Denoting the rate of a user's tweet activity as $\rho = (\beta(\langle k \rangle + 1) + \langle \lambda_0 \rangle \langle k \rangle M) / M$, and following the analysis from Ref.~\citenum{gleeson2014competition}, the differential equation for $G(t,x)$ is obtained (see Appendix \ref{si:2}):
\begin{equation}\frac{\partial G}{\partial t} = \lambda_0 x f(G) + \rho - (\lambda_0 + \rho)G,	\label{G} \end{equation}
which can be solved with initial conditions $f^\prime(1) = \langle k \rangle$ and $G(0,x) = 1$.

The above PGF provides a prediction of the expected popularity $m(t)$ for the focal tweet at time $t$, and by definition the number of retweets is $m(t) - 1$.
In the case of constant infectivity with no decay effect, equation (\ref{G}) leads to (see Appendix \ref{si:2})
\begin{equation}
	m(t) = (2\lambda_0 + \rho) \tau + (1 - (2\lambda_0 + \rho)\tau)\exp\left({- {t}/{\tau}}\right),	\label{mt}
\end{equation}
where $\tau \equiv 1 / (\rho - \lambda_0 (\langle k \rangle - 1))$.
When $\lambda_0$ is small enough such that $\tau > 0$, equation (\ref{mt}) suggests that the popularity converges to a finite level. In contrast, for $\lambda_0$ large enough and $\tau < 0$, equation (\ref{mt}) indicates that popularity grows exponentially with time.
The threshold separating these two behaviours is at
\begin{equation}
	\bar{\lambda_0} = \rho / (\langle k \rangle - 1). \label{threshold}
\end{equation}
Above this threshold, information can spread to a global scale; However, when $t \to \infty$ the exponential growth prediction $m(t) \to \infty$ does not conform with real data, calling for additional effects to reproduce the empirical process.

\subsection{Decay factor and infectivity estimation.}
Previous studies have found that the attractiveness of online information does not remain constant over an indefinite period of time, but rather gradually declines as it grows older \cite{wu2007novelty}.
We adopt this observation of fading popularity by incorporating a decay factor $\alpha$ and assume that the infectivity of cascade $i$ decays exponentially by $\lambda_{i}(t)=\lambda_{i0}e^{-\alpha(t-t_{i0})}$, where $t_{i0}$ is the time of the initial tweet.
Among retweets for which we can identify at least one of the previous tweets in the same cascade posted by their neighbours, a fraction $\psi = 0.69$ of them occurred within one day after the tweet was last seen by the retweeted user.
Using a mean-field approach that assumes the degree of all nodes to be equal to $\langle k \rangle$, we then express the average number of retweets of cascade $i$ at time $t$ as $a_{i,t} = \lambda_{i0} e^{-\alpha t} \langle k \rangle a_{i, t-1} / \psi$.

We define the number of total retweets of cascade $i$ at time $t$ as $A_{i, t}$, and derive the conditional expectation of $A_{i, t}$ given that cascade $i$ is retweeted at least once during its lifetime:
\begin{equation} E(A_{i, t} | A_{i, t} \geq 1) \equiv \sum_{\tau = 1}^{t} a_{i, \tau} = \sum_{\tau = 1}^{t} (\frac{\lambda_{i0} \langle k \rangle}{\psi})^{\tau} e^{-\frac{1}{2}\alpha \tau(\tau + 1)}. \label{A_t}
\end{equation}
Here we make two assumptions about the retweet size and infectivity of cascades: first, the tweet will either be stifled by stochastic fluctuations at the beginning such that no followers retweet it, or will be retweeted with probability $\langle k \rangle \lambda_{i}(t) \psi^{-1}$ and reach the mean size determined by equation (\ref{A_t}) at time $t$;
second, for fixed values of $t$ and $A_{i, t}$, the infectivity $\lambda_{i0}$ calculated by equation (\ref{A_t}) is the minimum rate to reach a retweet size $\geq A_{i, t}$.
We further assume that the relation between the number of retweets $S_i$ in the real Twitter data and $A_{i, t}$ is $S_i = A_{i, t}|_{t \to \infty}$. Then we set $t = 25$ to fit the spread rate distribution in equation (\ref{A_t}).
As such, we can obtain $(\lambda_{i0}, S_i)$ pairs such that their probability distribution satisfies $P(S \geq S_i) = P(\lambda_0 \geq \lambda_{i0})$, which can be used to approximately estimate the distribution of $\lambda_0$ from empirical Twitter data (Appendix \ref{si:3}).

The above analysis has taken the decay effect into account. We next approximate the distribution of initial infectivity $\lambda_{i0}$ for cascade $i$ as a truncated lognormal form with an upper bound probability $\lambda_{\mathrm{max}}$.
Let $p^0(\lambda_0)$ be the lognormal distribution $p^0(\lambda_0) = (\lambda_0 \sigma \sqrt{2 \pi}) ^ {-1} e^{-(\ln \lambda_0 - \mu)^2 / 2\sigma^2}$, where $\mu$ and $\sigma$ are parameters, and the normalization factor for the infectivity distribution can be written as $P^0(\lambda_{\mathrm{max}}) = \int_{0}^{\lambda_\mathrm{max}}p^0(\lambda_0)\mathrm{d}\lambda_0$.
Thus we have the probability distribution of infectivity $p^{\mathrm{infectivity}}(\lambda_0) = p^0(\lambda_0) / P^0(\lambda_{\mathrm{max}})$ in the truncated lognormal form with $0 < \lambda_0 < \lambda_{\mathrm{max}}$.
If a random user tweets a cascade with initial infectivity $\lambda_0$, and it stays in the followers' memory for an average lifetime $1 / \psi$, the probability that it is not retweeted by any follower is $(1-\lambda_0) ^ { \langle k \rangle / \psi}$.
Therefore, the fraction of cascades being retweeted at least once is given by
\begin{equation} P(\lambda_0) = \int_{0}^{\lambda_0} p^{\mathrm{infectivity}}(\tau)[1-(1-\tau)^{\langle k \rangle / \psi}]\mathrm{d}\tau.
	\label{p_ret} \end{equation}
This expression captures the fact that information cascades are likely to be stifled due to stochastic fluctuations at the initial stage, before it actually starts spreading.
Assuming the infectivity is small such that $[1-(1-\lambda_0)^{\langle k \rangle / \psi}] \simeq \lambda_0 \langle k \rangle / \psi$,
we have \begin{align}P(\lambda_0) & = \frac{\langle k \rangle}{2 \psi P^0(\lambda_{\mathrm{max}})} e^{\mu + \frac{\sigma^2}{2}} \Big [ 1+\mathrm{erf}(\frac{\ln\lambda_0 - \mu - \sigma^2}{\sigma\sqrt{2}}) \Big ],
	\label{p_theory} \end{align}
where $\mathrm{erf}(x)$ is the error function.
We then estimate $(\lambda_{i0}, S_i)$ pairs from empirical data with a pre-assumed decay factor $\alpha$ from equation \ref{A_t}, and fit the outcome distribution with equation \ref{p_theory} (see Fig.~\ref{main_results}a).

\subsection{Simulations.}
The simulations start with a set of users generating tweets, the infectivity of which follow a truncated lognormal probability distribution, with a universal decay factor governing their long time dynamics.
When a user tweets a new message by herself, or retweets an old message from her followees, illustrated in Fig.~\ref{Twitter}, all of her followers will receive the message.
A user will only see the latest tweets within her memory length, which is a fixed value for all users \cite{weng2012competition, gleeson2016effects, qiu2017limited,sreenivasan2017information}.
A natural measure of popularity is the number of retweets plus one that accounts for the original tweet, and we regard each not-retweeted tweet as a cascade with popularity $1$.
The innovation probability (the probability that a user generates a brand new tweet) $\beta = 0.528$ is calculated from Twitter data.

The mean degree of the Twitter follower network is $\langle k \rangle = 47.94$ with a total number of $N_{cas} = N_{twt} - N_{ret} = 10,366,501$ cascades, of which $962,341$ are cascades with popularity $> 1$.
Each time step a user retweets or creates on average $N_{twt}/NT$ cascades that will be retweeted $\langle \lambda \rangle \langle k \rangle N_{twt}/NT$ times by her followers in the next time step, leading to an estimate of average infectivity as $\langle \lambda \rangle = \psi N_{ret} / \langle k \rangle N_{twt} = 0.002$. Memory length can thus be estimated by $M = \psi N_{ret} / \langle \lambda \rangle N T = 43$, and the threshold $\bar{\lambda}$ in equation (\ref{threshold}) is $0.015$.
We use decay parameter $\alpha = 0.01$, and the corresponding infectivity distribution parameterized by $\mu = \ln 0.0012$, $\sigma = \ln 2.4$ and $\lambda_{\mathrm{max}} = 0.017$ to obtain the blue curve in Fig.~\ref{main_results}a fitting to the red dots of $(\lambda_{i0}, S_i)$ pairs calculated from Twitter data.

In all simulations, we first run a burn-in period of $100$ time steps. As the Twitter data focus on new memes (see Appendix \ref{si:1}), we only analyse new cascades that originate in the next $T = 33$ time steps (see Appendix \ref{si:4} for more details).

\section{Empirical Twitter data}
\label{si:1}
The empirical data we use in this paper, developed by and studied previously in Refs.~\citenum{weng2013virality} and \citenum{weng2014predicting}, were sampled from Twitter between March 24, 2012 and April 25, 2012. 
New memes are defined as those with fewer than 20 tweets during the previous month, and only new memes that emerged during the observation time window were selected. 
The data contain the follower network with reciprocal following ties, the timeline hashtag data set of tweets generated by the users in the follower network, and the timeline hashtag data set of retweets with information of the retweeted user along with the user from whom the tweet originated. 

\begin{figure*}[!htbp]
	\centering \includegraphics[width=0.92\linewidth]{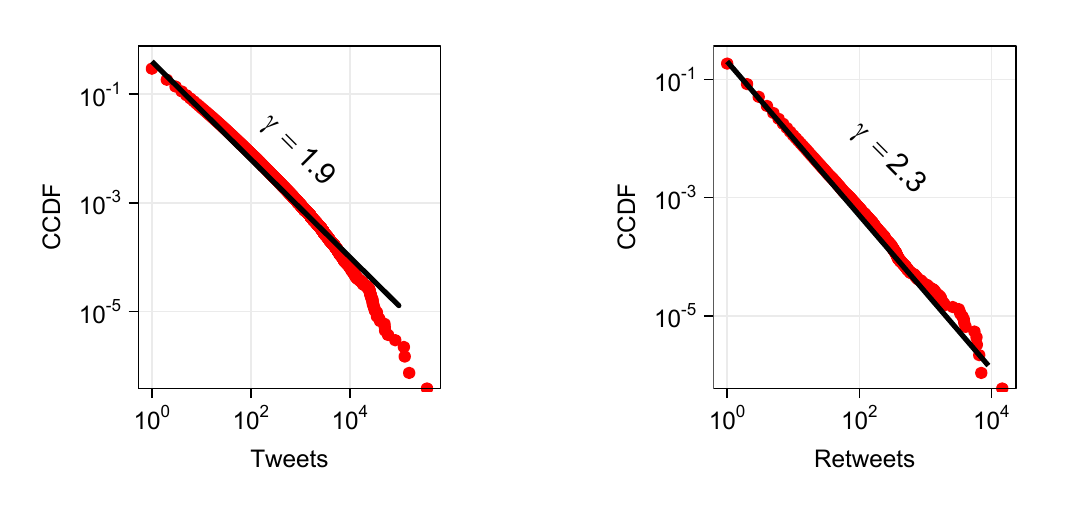} 
	\caption{\textbf{Distribution of tweets and retweets in Twitter data.} \emph{(Left)}: complementary cumulative probability distribution (CCDF) of hashtags (memes); \emph{(Right)}: CCDF of cascades. Black lines indicate power laws with given exponents $\gamma$.}
	\label{distribution}
\end{figure*}

Many studies have used hashtags as memes for exploring information diffusion processes in online social platforms \cite{weng2012competition, gleeson2014competition}.
Displaying distribution statistics from empirical Twitter data for both hashtags (or memes) and retweet cascades in \ref{distribution}, we show that they are likely to be driven by different underlying dynamics.
For example, under fitting a power-law distribution (without arguing about the validity of doing so) the exponents are fairly different: $\gamma = 1.9$ for hashtags and $\gamma = 2.3$ for cascades.
Distributions with $\gamma < 2$ and $2 < \gamma < 3$ exhibit distinct statistical features: the first moment of a power-law distribution with $\gamma < 2$ is infinite, while it is finite for distributions with $\gamma > 2$, indicating that on average a hashtag appears in a huge number of tweets, while the size of a cascade is usually moderate.
Viral hashtags can reach broader audience, have more extensive global impact, and are largely influenced by broadcasting. In contrast, retweet cascades are mostly spread by more immediate followers, have a more local impact, and are mainly transmitted via link contagion through personal ties in social networks.

\section{Analysis of branching dynamics}
\label{si:2}
We adopt the underlying network topology of Ref.~\citenum{gleeson2014competition}, in the form of a directed social network such as Twitter, where there are $N$ users represented by nodes in the system. In our analysis we will regard $N \to \infty$. A randomly picked node has $k$ followers with probability $p_{k}$, while it follows $\langle k \rangle$ others with $\langle k \rangle$ denoting the mean out-degree $\langle k \rangle=\sum_{k}kp_{k}$. The out-degree follows the power-law distribution with $p_{k} \propto k^{-\gamma}$. We set up an identical user memory length $M$ for every individual. Only tweets within the $M$th ranking can be seen and retweeted by the user. The ranking only accounts for the aging of tweets with new tweets always ranking higher than old ones.

We set the infectivity of the focal tweet as $\lambda_0$ and the average infectivity as $\langle \lambda_0 \rangle$ for all other tweets, which is the probability that a tweet will be chosen to be retweeted. 
For simplicity, the decay effect of tweets will not be considered initially here (but will be introduced further below).
To measure the activity of tweet creation by users, we define the innovation rate as $\beta$, which is the probability that a user generates a brand new tweet. Throughout this section we consider a small time window $\Delta t$ so that for any user at most one tweet will be created during this time. A user can also retweet old tweets on the screen with probabilities according to their infectivities. All of these tweets, whether innovated or retweeted, will appear on the screen of its followers, by overwriting existing tweets. To simplify the analysis, we assume that during updates any of the old existing tweets will be overwritten with the same probability. Thus if a user receives $l$ new tweets from her followed nodes, we randomly pick $l$ slots on her screen to clear out to write the new ones on. 

If the focal tweet is tweeted by the user, then it will be cleared from her screen. The vacant slot will be filled in by a randomly chosen tweet whose infectivity is $\langle\lambda_0\rangle$. At any time, as a user has $\langle k \rangle$ followed nodes, there will be on average $(\langle k \rangle+1)\beta$ newly generated tweets and $\langle \lambda_0 \rangle\langle k \rangle M $ retweeted existing tweets in the next time step. If a user knows the focal tweet already, in other words she has the tweet on the screen, then the probability that the focal tweet will be overwritten in the next time step is

\begin{equation}
	\rho=\left\{
	\begin{array}{rl}
		&1,\qquad \qquad \qquad \frac{\beta(\langle k \rangle+1)+\langle \lambda_0 \rangle\langle k \rangle M }{M} \geq 1,
		\\
		&\frac{\beta(\langle k \rangle+1)+\langle \lambda_0 \rangle\langle k \rangle M }{M},\qquad \frac{\beta(\langle k \rangle+1)+\langle \lambda_0 \rangle\langle k \rangle M }{M} < 1.
	\end{array} \right.
\end{equation}
Throughout this paper we set parameters so that
$\frac{\beta(\langle k \rangle+1)+\langle \lambda_0 \rangle \langle k \rangle M }{M} < 1$. Thus the overwriting probability is given by
\begin{equation}
	\rho=\frac{\beta(\langle k \rangle+1)+\langle \lambda_0 \rangle   \langle k \rangle M}{M}.
\end{equation}

The tweet we are interested in starts to spread from a randomly chosen root screen.
Each retweet of the tweet adds $1$ to the popularity.
We denote $G(a,x)$ as the probability generating function (PGF) of the excess popularity distribution.
At age $a$ (i.e., at time $t+a$, where $t$ is the birth time for the focal tweet), we define the PGF $H(a,x)$ for the popularity distribution of our focal tweet \cite{gleeson2014competition}
\begin{equation}
	H(a,x)=\sum_{n}^{}q_{n}(a)x^n,
\end{equation}
where $q_{n}(a)$ is the probability that the tweet has been retweeted $n$ times. It is also convenient to define the PGF $G^{(k)}(a,x)$ for the popularity distribution that the focal tweet originates from a root screen with out-degree $k$ \cite{newman2001random, wilf2013generatingfunctionology}
\begin{equation}
	G(a,x)=\sum_{k}p_{k}G^{(k)}(a,x),
\end{equation}
and 
\begin{equation}
	H(a,x)=xG(a,x)f(G(a,x)).
\end{equation}

For simplicity, we just focus on the model with only one node with the tweet on her screen at the initial stage.
Now consider the focal tweet posted up for the first time on a screen with out-degree $k$ (call this screen $S1$), at time $0$. We let $T(a)$ be the random variable for the number of tweets originated from a randomly picked node, and $T_{k}(a)$ the number of tweets originated from a node with degree $k$, at time $a$ for the focal tweet. In other words, the tweet has age $a$ at the observation time. At the next time step $\Delta t$, there will be three possible outcomes on screen $S1$ that contribute to the PGF $G^{(k)}(a,x)$:

(1) The tweet is retweeted, then removed from the screen $S1$, and shows up on the screens of all its $k$ followers. This happens with probability $\lambda_0 \Delta t$, and the number of tweets under this scenario, denoted by $T_{k,1}(a)$, can be further rewritten by $T_{k,1}(a)=1+kT(a-\Delta t)$, as the degree of the followers are random. Note that at time $\Delta t$, the age will be $a-\Delta t$ at the observation time. This contributes $x[G(a-\Delta t,x)]^{k}$ to $G^{(k)}(a,x)$.

(2) The tweet is not retweeted but overwritten by other tweets appearing on screen $S1$ during this time period. This happens with probability $(1-\lambda_0\Delta t)\rho \Delta t= \rho \Delta t+o((\Delta t)^2)$, and the number of tweets for this outcome is $T_{k,2}(a)=0$. This contributes $1$ to $G^{(k)}(a,x)$.

(3) The tweet doesn't retweet, and it survives this period of time. This happens with probability
$(1-\lambda_0\Delta t)(1-\rho \Delta t)=1-\lambda_0 \Delta t-\rho \Delta t+o((\Delta t)^2),$ and $T_{k,3}(a)=T_{k}(a-\Delta t)$. Thus it contributes $G^{(k)}(a-\Delta t,x)$ to $G^{(k)}(a,x)$.

Putting (1), (2) and (3) together, as each outcome is independent and exclusive to the others, we have
\begin{equation}
	\begin{split}
		G^{(k)}(a,x)=\lambda_0 \Delta t x[G(a-\Delta t,x)]^{k}+\rho \Delta t
		\\
		+(1-\lambda_0 \Delta t-\rho \Delta t)G^{(k)}(a-\Delta t,x),
	\end{split}
\end{equation}
which is correct to first order in $\Delta t$. Regarding $\Delta t \to 0$ we arrive at
\begin{equation}
	\frac{\partial G_{(k)}}{\partial a}=\lambda_0 x{[G]}^{k}+\rho-(\lambda_0+\rho )G^{(k)}.
	\label{G_k}
\end{equation}

Multiplying $p_{k}$ on both ends of equation (\ref{G_k}) and summing over all $k$ yields
\begin{equation}
	\frac{\partial G}{\partial a}=\lambda_0 x f(G)+\rho-(\lambda_0 +\rho)G.
	\label{G}
\end{equation}

We now use this partial differential equation to find the mean popularity of the focal tweet at age $a$
\begin{equation}
	m(a)\equiv \sum_{n=1}^{\infty}nq_{n}(a)=\frac{\partial H}{\partial x}(a,1)=1+(1+\langle k \rangle)\frac{\partial G}{\partial x}(a,1).
	\label{m_def}
\end{equation}
Note that $G(a,1)=1$, $f(1)=1$, $f^{\prime}(1)=\langle k \rangle$ and $m(0)=1$. Differentiating equation (\ref{m_def}) with respect to $x$ we have
\begin{equation}
	\frac{\mathrm{d}m}{\mathrm{d}a}=\lambda_0 (1+\langle k \rangle)+(\lambda_0 \langle k \rangle-\lambda_0 -\rho)(m-1),
	\label{m1}
\end{equation}
with $m(0)=1$. We now make further observations about the infectivity $\lambda_0 $ of the focal tweet.

The focal tweet has a constant infectivity during the entire diffusion process. We use $\tau=1 / (\rho-\lambda_0(\langle k \rangle-1))$ to rewrite equation (\ref{m1}) to obtain the following result:

\begin{equation}
	m(a) = \left\{
	\begin{array}{rl}
		(2\lambda_0 + \rho) \tau + (1-(2\lambda_0+\rho)\tau)\exp\left({-\frac{a}{\tau}}\right),\quad &\lambda_0 \neq \frac{\rho}{\langle k \rangle-1},
		\\
		1+\lambda_0(1+\langle k \rangle)a,\qquad \qquad \qquad \qquad \qquad \quad &\lambda_0 = \frac{\rho}{\langle k \rangle-1}.
	\end{array} \right.
\end{equation}

Here a spreading threshold appears with
\begin{equation}
	\bar{\lambda_0}=\frac{\rho}{\langle k \rangle-1}=\frac{\beta(\langle k \rangle+1)+\langle\lambda_0 \rangle \langle k \rangle M}{M(\langle k \rangle-1)}.
	\label{threshold}
\end{equation}
It shows that tweets with infectivity less than $\bar{\lambda_0}$ typically won't successfully spread out; they are likely to be forgotten before being retweeted even once.


\section{Infectivity distribution}
\label{si:3}
We propose a simple method to estimate the infectivity distribution of cascades from Twitter data.
For a given retweet size $S_i$, the fraction of cascades with size $\geq S_i$ is calculated by $P(S \geq S_i) = N(S \geq S_i) / N_{cas}$, where $N(S \geq S_i)$ is the number of cascades with size $\geq S_i$. 
To associate the corresponding $\lambda_{i0}$ with $S_i$, for a given decay factor $\alpha$, we set the total time $t = 25$ in equation (4) in the main text and let $E(A_{i,t}|A_{i,t} \geq 1) = S_i$ to calculate the $\lambda_{i0}$ on the right side. 
We assume that to reach a cascade of at least $S_i$ retweets, the minimum infectivity is $\lambda_{i0}$ calculated above, thus we could derive a complementary cumulative probability distribution of infectivity from Twitter data by $P(\lambda_0 \geq \lambda_{i0}) = P(S \geq S_i) = N(S \geq S_i) / N_{cas}$, which are the dot plots in red and orange in Fig. 4a.
Meanwhile, we assume that a cascade with infectivity $\lambda_{i0}$ will either not be retweeted at all due to initial fluctuations, or will reach a retweet size $S_i$ determined by equation (4) in the main text.
With a given set of $\mu$, $\sigma$ and $\lambda_{\mathrm{max}}$ in the truncated lognormal distribution, we could fit a complementary cumulative probability distribution by $P(\lambda_0 \geq \lambda_{i0}) = P(\lambda_{\mathrm{max}}) - P(\lambda_{i0})$, which is the blue curve in Fig. 4a.
Therefore we can compare theoretical lognormal distribution to the corresponding distribution of Twitter retweet data, with our parameter setting matching the real data quite well.

\begin{figure*}[!htbp]
	\centering 
	\includegraphics[width=0.92\linewidth]{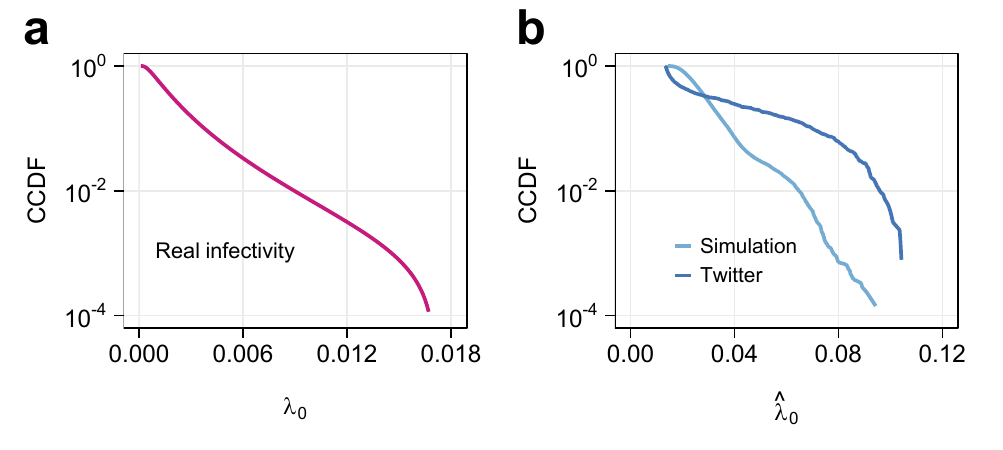}
	\caption{\textbf{Real and estimated infectivity distributions.} \textbf{a}, Real infectivity distribution used for simulation models with $\mu = \log 0.0012$, $\sigma = \log 2.4$, $\lambda_{\mathrm{max}} = 0.017$. \textbf{b}, Estimated infectivity based on the first $50$ retweets in simulation and Twitter retweet data.}
	\label{fig::infectivity_distr}
\end{figure*}

Note that the lognormal parameters change if we reset the decay parameter $\alpha$.
In the main text we use $\alpha = 0.01$ with $\mu = \ln 0.0012$, $\sigma = \ln 2.4$ and $\lambda_{\mathrm{max}} = 0.017$. 
We show the true infectivity distribution used in simulations and the estimated infectivity distribution for cascades with at least $50$ retweets in  \ref{fig::infectivity_distr}.
The estimated infectivities of popular cascades is usually larger than the $\lambda_{\mathrm{max}}$ used in simulations. This suggests that our method overestimates the infectivities of popular cascades while underestimating that of not-retweeted cascades by an infectivity $\hat{\lambda_0} = 0$.

\section{Simulation details}
\label{si:4}
In our simulation model, we use one day as one time step, since our model requires that the rate of user activity is homogeneous across time steps (e.g., most people don't use Twitter after midnight). 
The simulation starts with each individual generating a cascade by innovation probability $\beta = 0.528$ at each time step, the infectivity of which is sampled from the truncated lognormal distribution $p^{\mathrm{infectivity}}(\lambda_0)$
where $0 <\lambda_0 <\lambda_{\mathrm{max}}$. The cascade will then be seen by all of her followers. The attention length of each user (the maximum number of tweets within her attention) is set to $M=43$. When new tweets appear, the oldest ones will be forgotten by the user. At each time step, a user can either post new cascades, or retweet any cascade received from others according to its infectivity. The infectivity of cascade $i$ will decrease according to the imposed decay factor as 
\begin{equation}
	\lambda_{i}(t)=\lambda_{i0}e^{-\alpha(t-t_{i0})}.
\end{equation} In the beginning of the simulation, no retweets are in the system. After around 10 time steps the number of retweets generated by the users within each time step will become stable. Therefore we take the first $100$ time steps of the simulation as the ``burn--in'' stage, and collect simulated retweet data from the $101$st to the $133$rd time step.

\section{Other model specifications}
\label{si:5}
We present simulation results based on other parameter specifications, as a robustness test for models discussed in the main text.
First, fixing values of other parameters, we change the decay factor $\alpha$ to new values $0.001$ and $0.02$, and the $\lambda_{\mathrm{max}}$ in lognormal distribution to $0.0158$ and $0.018$, respectively.
The fit to equation (6) in the main text is shown in \ref{rateSI}, suggesting that when changing the decay parameter we can still fit well to the lognormal distribution of infectivities of cascades.

\begin{figure*}[!htbp]
	\centering \includegraphics[width=0.6\linewidth]{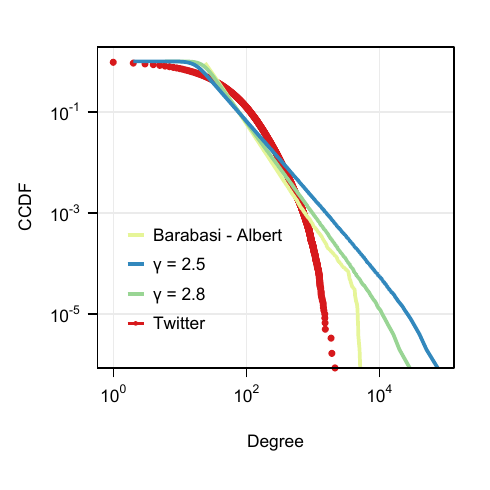} 	
	\caption{\textbf{Degree distribution of networks.} The degree distribution of the undirected Twitter follower network of reciprocal ties, and other synthetic networks used in our simulations. We later show that when a synthetic network has a similar degree distribution to the Twitter network, we could reproduce the cascade size distribution of retweets.}
	\label{deg_distribution}
\end{figure*}

\begin{figure*}[!htbp]
	\centering 
	\includegraphics[width=0.92\linewidth]{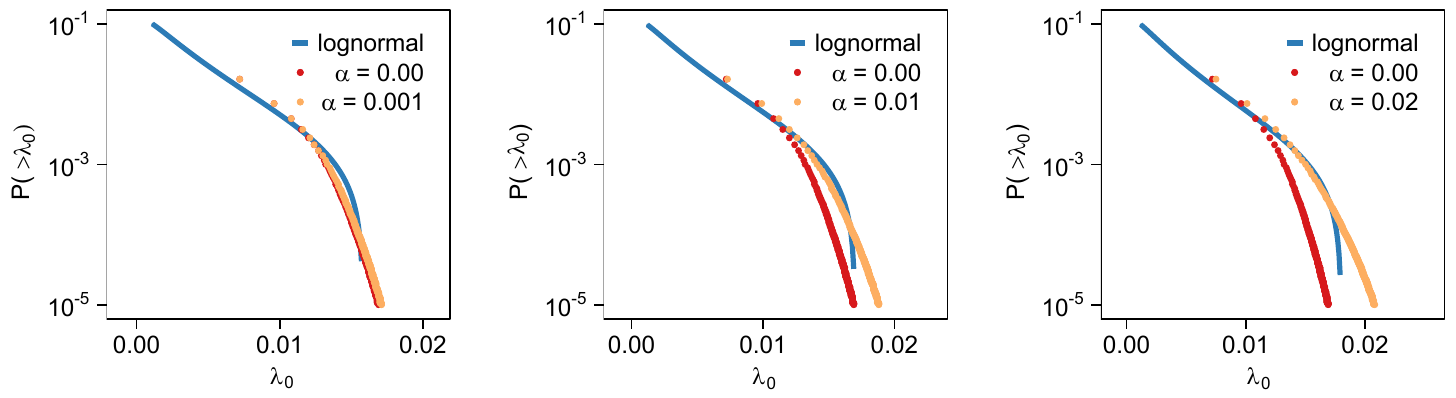}
	\caption{\textbf{Lognormal distribution fit with different decay parameters}. (\emph{Left}): $\alpha = 0.001$ and $\lambda_{\mathrm{max}} = 0.0158$; (\emph{Middle}): $\alpha = 0.01$ and $\lambda_{\mathrm{max}} = 0.017$; (\emph{Right}): $\alpha = 0.02$ and $\lambda_{\mathrm{max}} = 0.018$. For all panels the other parameters are the same as those in the main text: $\mu = \log 0.0012$ and $\sigma = \log 2.4$.}
	\label{rateSI}
\end{figure*}

We try to replicate our simulation model in several networks.
In \ref{deg_distribution} we show the degree distribution of some of the candidate networks, including the Twitter reciprocal network, the Barab{\`a}si-Albert network \cite{barabasi1999emergence}, and synthetic networks \cite{goh2001universal} with exponents $\gamma = 2.5$ and $\gamma = 2.8$. More detailed statistics of these networks are presented in \ref{net_stat}. 

\begin{table*}[!htbp]
	\centering
	\caption{\textbf{Statistics of networks used in simulation models.} We present here the detailed network statistics of the Twitter follower network, the Barabasi - Albert network, synthetic networks in Ref.~\citenum{goh2001universal} with exponent $\gamma = 2.8$ and $\gamma = 2.5$. }
	\begin{tabular}{cccccc}\\
		\hline \hline
		{Network} & {{N}} & $\langle k\rangle$ & $\langle k^2 \rangle$  & $k_{\mathrm{max}}$ & $\langle k ^2 \rangle / \langle k \rangle$ \\
		\hline
		Twitter	& $5.95 \times 10^5$&$47.94$	& $7.29 \times 10^3$	& $2.15 \times 10^3$ & $1.52 \times 10^2$\\		
		Barabasi - Albert	& $5 \times 10^5$ &$48$	& $7 \times 10^3$	& $5 \times 10^3$ & $1.5 \times 10^2$ \\
		Synthetic network $\gamma = 2.8$	& $5 \times 10^5$&$48$	& $1 \times 10^4$	& $3 \times 10^4$ & $3 \times 10^2$\\
		Synthetic network $\gamma = 2.5$	& $5 \times 10^5$&$48$	& $5 \times 10^4$	& $7 \times 10^4$ & $1 \times 10^3$\\	
		\hline \hline
	\end{tabular}
	\label{net_stat}
\end{table*}	

To test whether the decay effect of infectivity affects the results of our model, we also run simulations with different parameter settings: When $\alpha = 0.001$, we fix $\mu = \ln 0.0012$, $\sigma = \ln 2.4$ and change $\lambda_{\mathrm{max}}$ to $0.0158$; When $\alpha = 0.02$, we also fix $\mu = \ln 0.0012$, $\sigma = \ln 2.4$ but change $\lambda_{\mathrm{max}}$ to $0.018$. 

\begin{figure*}[!htbp]
	\centering	\includegraphics[width=0.98\linewidth]{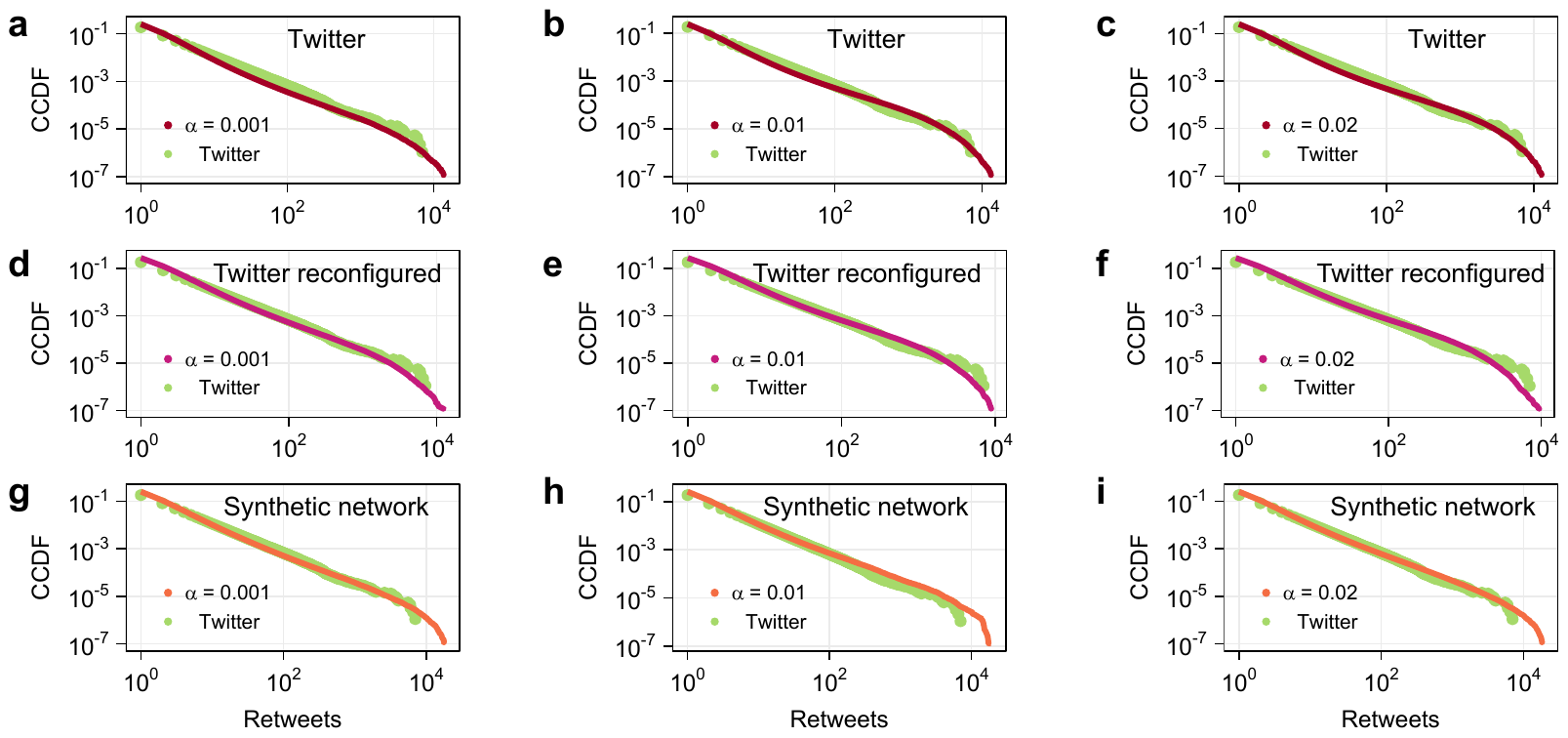}
	\caption{\textbf{Simulation on Twitter follower network and other synthetic networks.} Parameter settings are:  (\emph{Left panel}): $\alpha = 0.001, \lambda_{\mathrm{max}} = 0.0158$; 	
		(\emph{Middle panel}): $\alpha = 0.01, \lambda_{\mathrm{max}} = 0.017$; 	
		(\emph{Right panel}): $\alpha = 0.02 \lambda_{\mathrm{max}} = 0.018$.
		Other parameters: $\mu = \ln 0.0012, \sigma = \ln 2.4, M = 43, \beta = 0.528$. For the synthetic networks, the power-law exponent is $\gamma = 2.8$ and mean degree $\langle k \rangle = 48$. We ran $1,000$ simulations for each plot.}
	\label{simSI}
\end{figure*}

With the above new parameter settings, we run simulations on the Twitter network, reconfigured random network that preserves the degree distribution of the empirical Twitter network, and a synthetic network \cite{goh2001universal} with power-law exponent $\gamma = 2.8$ to verify results in the main text.
The selected networks appear to be reasonable power-law approximations to the degree distribution of the empirical Twitter network, and the simulation results in these networks match the empirical retweet distribution well (\ref{simSI}).
These findings show that our model is not sensitive to specific network topology with similar degrees of heterogeneity.

\section{Prediction and random forests models}
\label{si:6}

Since our simulation model does not allow an individual to retweet the same cascade repeatedly, the ``user entropy'' from Ref.~\citenum{weng2013virality} is the same as retweet entropy $H^r$ and therefore not discussed in this paper. In addition, Weng et al. defined four baseline models of information spread in Ref.~\citenum{weng2013virality}: the random sampling model (M1); the simple cascade model (M2) that accounts only for the network structure; the social reinforcement model (M3) that chooses the user with maximum number of infected neighbours to adopt the cascade; and the homophily model (M4), which assumes that only neighbours in the same community can retweet the cascade. 

The detailed definition of baseline models are as follows: For a given cascade, M1 randomly samples the same number of retweets as in the real data.
M2 randomly selects a user, and at each time step with probability $0.85$, we randomly select one of its neighbours to retweet, or with probability $0.15$ the process restarts from a new user. Comparing to M1, M2 takes the network structure into account. 
The cascade in M3 is generated similarly to M2 but at each time step the user with the maximum number of infected neighbours retweet the cascade. M3 accounts for the social reinforcement effect.
M4 simulates in the same way as in M2 but at each step, only neighbours in the same community can retweet the cascade, which accounts for homophily effect \cite{weng2013virality}. 

\begin{figure}[!htbp]
	\centering
	\includegraphics[width=0.98\linewidth]{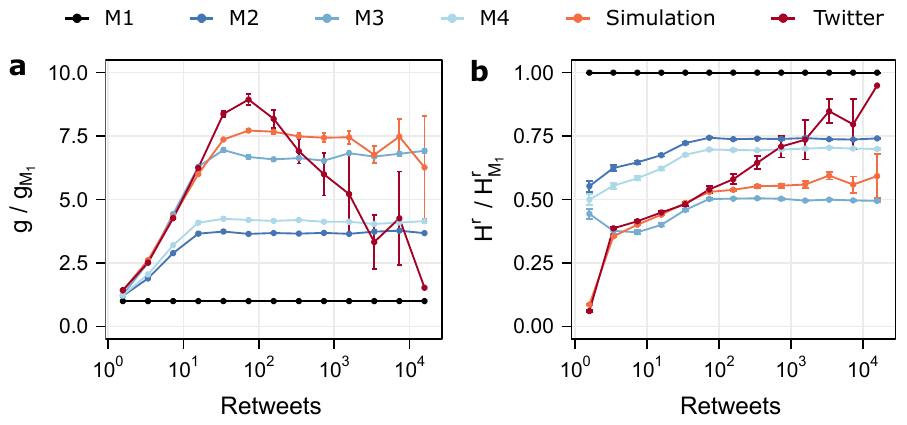}
	\caption{\textbf{Statistics based on community structure.} \textbf{a}, Adoption dominance $g$ for baseline models M1, M2, M3, M4, Twitter retweet data and the simulation retweet data of attention limit model based on the first $50$ retweets. \textbf{b}, Retweet entropy $H^r$ for baseline models M1, M2, M3, M4, Twitter retweet data and the simulation retweet data of attention limit model based on the first $50$ retweets. Our simulation model is closest to real Twitter data statistics.
	}
	\label{Hu_gu}
\end{figure}

As a direct comparison between our simulation model and the baseline models, in  \ref{Hu_gu} we present the $g$ and $H^r$ statistics of cascade diffusion scaled by that of M1 for the other baseline models and our simulation model. Though our model assumptions do not account for community structures, it nevertheless outperforms other baseline models.

We use the \emph{InfoMap} algorithm for community detection in the Twitter follower network \cite{rosvall2008maps}. We run random forests models by using 10-fold cross validation.
Variables used in random forests models include: 

Two null model predictors:  The number of distinct users: the number of distinct retweet users in the first $50$ retweets of a given cascade; 
The total number of neighbours of early retweet users: for each cascade, sum up the number of users who retweeted the first $50$ retweets \cite{weng2013virality};

Three community--based predictors: The number of infected communities: the number of distinct communities that has at least one user who retweeted in the first 50 retweets;
Retweet entropy $H^r$: the entropy based on how retweet users of a cascade are distributed across different communities. It is computed based on the first 50 retweets;
Fraction of intra--community user interactions: count pair-wise user interactions for a given cascade, and compute the proportion that occur between users in the same community. For retweet data, as we only know the original user who posted the retweet and the users who retweeted it, this measure becomes the fraction of first $50$ retweet users who are in the same community as the original user \cite{weng2013virality};

Estimated infectivity $\hat{\lambda_0}$: we use equation (2) in the main text to estimate the infectivity of a given cascade based on the first 50 retweets. Note that this should be interpreted as the infectivity without a decay parameter. To numerically obtain $\hat{\lambda_0}$ we start from $\hat{\lambda_0}=0$ and increase by $0.0001$ to get the value that makes the right side of equation (2) closest to $50$.

The random forests results are presented in Fig.~3 and \ref{table:rf}. Note that in \ref{table:rf} we also run models using the true infectivity $\lambda_0$ on simulation models. This improves the prediction and recall rates substantially compared to the random forests models using the estimated infectivity $\hat{\lambda_0}$, suggesting that an improved estimation method for infectivity could be a key factor for the improved prediction of viral cascades in Twitter.

We also run logistic regressions with the predictors used in random forests models in \ref{table:logit}, \ref{table:logitSimLambar}, and \ref{table:logitSimLam}. 
These results are discussed in detail in the main text of the paper.

\begin{table}[!htbp]
	\centering
	\caption{\textbf{Random forests results in 10-fold cross validation.} Cascades with size $S \geq 50$ are considered. Simulated retweet data are generated with $\alpha =0.01$ from $20$ replications. All models in this table use the two null model predictors: The number of distinct users, and the total number of neighbours of early retweet users. There are three community--based predictors: The number of infected communities, user entropy $H^u$, and the fraction of intra--community user interactions. $\hat{\lambda_0}$ is the infectivity estimated by equation (2) without accounting for the decay factor. $\lambda_0$ is the real infectivity used in simulations. 
	}
	\hspace*{-.8cm}	\begin{tabular}{|c|c|c|c | c| c| c|c| c|c| c|}
		\hline \hline 
		\multirow{2}{*}{Model} & \multirow{2}{*}{Data} & \multirow{2}{*}{$\hat{\lambda_0}$} &\multirow{2}{*}{$\lambda_0$} &\multirow{2}{*}{Community}& \multicolumn{2}{|c|}{$\theta = 90\%$} & \multicolumn{2}{|c|}{$\theta = 80 \%$} & \multicolumn{2}{|c|}{$\theta = 70 \%$}\\
		\cline{6-11} 
		& &  & & & Precision & Recall & Precision & Recall & Precision & Recall\\
		\hline 
		NM		& Simulated & &  & &  0.23  &  0.15 & 0.34 &  0.29 & 0.42 & 0.39\\
		CB		& Simulated &  & & $\surd$ &	 0.42 & 0.11 & 0.51  & 0.23 & 0.54 & 0.35 \\
		IB		& Simulated &$\surd$ & & &    0.42  & 0.24 & 0.53  & 0.40 & 0.61 &  0.53\\
		C\&I	& Simulated &$\surd$ & & $\surd$ &  0.57  & 0.24 & 0.66 & 0.40 & 0.69 & 0.55\\
		IB ($\lambda_0$)	&	Simu & & $\surd$ & &     0.59 & 0.43 & 0.65 & 0.53 &	0.68 & 0.60\\
		C\&I ($\lambda_0$)	&	Simu & & $\surd$ & $\surd$ &   0.65 & 0.42 & 0.71 & 0.54 & 0.73 & 0.61\\
		
		\hline 
		NM	&	Twitter    & & 	& & 0.15 & 0.10 & 0.32 & 0.26 & 0.40 & 0.33\\
		CB	&	Twitter    & & &$\surd$ & 0.52 & 0.15 & 0.54 & 0.29 & 0.55 & 0.34\\
		IB	&	Twitter    &$\surd$	& & & 0.38 & 0.22 & 0.55 & 0.44 & 0.59 & 0.51\\				
		C\&I	&	Twitter	   &$\surd$ & &$\surd$ & 0.59 & 0.21 & 0.65 & 0.44 & 0.70 & 0.53\\
		\hline \hline 
	\end{tabular}
	\label{table:rf}
\end{table}

\begin{table}[!htbp]
	\centering
	\caption{{\textbf{Logistic models of viral cascade prediction in Twitter data with estimated infectivity $\hat{\lambda_0}$.}} 
		In all columns, variables such that $p < .05$ are highlighted with one asterisk, while variables such that $p < .01$ are highlighted with two asterisks. Standard errors are shown in parentheses.} 
	\hspace{-1cm}
	\begin{tabular}{|c|c|c|c|}
		\hline 
		\hline  
		Dependent	& $\theta = 90\%$ & $\theta = 80 \%$ & $\theta = 70 \%$	\\
		
		\hline 
		Intercept		              &  ${-5.3}^{**}$ $(1.2)$ &  ${-4.0}^{**}$ $(0.85)$ & ${-3.1}^{**}$ $(0.70)$    \\
		$\hat{\lambda_0}$		      &  $-0.58$ $(4.7)$ &  $5.43$ $(3.6)$ & ${8.0}^{*}$ $(3.2)$   \\
		Early Adopters		          &  $0.03$ $(0.02)$ &  $-1.8 \times 10 ^ {-3}$ $(0.01)$ & $-3.4 \times 10 ^ {-3}$ $(0.01)$ \\
		Neighbours		              &  $6.4 \times 10 ^ {-6}$ $(6.5 \times 10 ^ {-5})$ &  $3.2 \times 10 ^ {-5}$ $(4.9 \times 10 ^ {-5})$& $1.2 \times 10 ^ {-5}$  $(4.3 \times 10 ^ {-5})$  \\
		Infected Communities		  &  ${-0.14}^{**}$ $(0.04)$ &  ${-0.12}^{**}$ $(0.03)$ & ${-0.11}^{**}$ $(0.02)$   \\
		$H^r$		                  &  ${2.1}^{**}$ $(0.63)$ &  ${2.1}^{**}$ $(0.48)$ & ${1.9}^{**}$  $(0.40)$  \\
		Intra-community  &  $0.80$ $(1.1)$ &   ${1.7}^{*}$ $(0.83)$ & ${1.7}^{*}$ $(0.69)$    \\
		\hline \hline 
	\end{tabular}
	\label{table:logit}
\end{table}

\begin{table}[t]
	\centering
	\caption{{\textbf{Logistic models of viral cascade prediction in simulation data with estimated infectivity $\hat{\lambda_0}$.}} 
		In all columns, variables such that $p < .05$ are highlighted with one asterisk, while variables such that $p < .01$ are highlighted with two asterisks. Standard errors are shown in parentheses. Note that as we do not allow one node to retweet the same cascade more than once, the number of early adopters for the first $50$ retweets is always $50$ and therefore not included in the logistic regressions.} 
	
	\hspace{-2cm}
	\begin{tabular}{|c|c|c|c|c|c|c|}
		\hline 
		\hline  
		Dependent		& $\theta = 90\%$ & $\theta = 80 \%$ & $\theta = 70 \%$\\
		
		\hline 
		Intercept		                  &  ${-5.4}^{**}$ $(0.17)$& ${-5.3}^{**}$ $(0.14)$ &  ${-5.5}^{**} $ $(0.13)$ \\
		$\hat{\lambda_0}$		        &  ${76.7}^{**}$ $(2.5)$ & ${1.1 \times 10 ^ {2}}^{**}$ $(2.6)$& ${1.5 \times 10 ^ {2}}^{**}$ $(2.9)$\\
		Neighbours		                 &  ${4.6 \times 10 ^ {-4}}^{**}$ $(1.5 \times 10 ^ {-5})$ & ${4.5 \times 10 ^ {-4}}^{**}$ $(1.3 \times 10 ^ {-5})$ & ${4.5 \times 10 ^ {-4}}^{**}$ $(1.2 \times 10 ^ {-5})$\\
		Infected Communities		      &  ${-0.04}^{**}$ $(0.01)$ & ${-0.04}^{**}$ $(0.01)$ & ${-0.05}^{**}$ $(0.01)$\\
		$H^r$		                   &  $-0.22$ $(0.13)$ & $-0.16$ $(0.10)$ & $-0.13$ $(0.09)$\\
		Intra-community  &  $-0.13$ $(0.15)$ & $-0.09$ $(0.12)$& $-0.15$ $(0.11)$\\
		\hline\hline  
	\end{tabular}
	\label{table:logitSimLambar}
\end{table}	

\begin{table}[t]
	\centering
	\caption{{\textbf{Logistic models of viral cascade prediction in simulation data with true infectivity ${\lambda_0}$.}} 
		In all columns, variables such that $p < .05$ are highlighted with one asterisk, while variables such that $p < .01$ are highlighted with two asterisks. Standard errors are shown in parentheses. Note that as we do not allow one node to retweet the same cascade more than once, the number of early adopters for the first $50$ retweets is always $50$ and therefore not included in the logistic regressions.} 
	
	\begin{tabular}{|c|c|c|c|c|c|c|}
		\hline 
		\hline  
		Dependent		& $\theta = 90\%$ & $\theta = 80 \%$ & $\theta = 70 \%$\\	
		\hline 
		Intercept		 &  ${-25.8}^{**}$ $(0.59)$ & ${-20.0}^{**}$ $(0.36)$ &  ${-16.4}^{**}$ $(0.28)$ \\
		$\lambda_0$ &  ${1.3 \times 10 ^ {3}}^{**}$ $(31.3)$ & ${1.00 \times 10 ^ {3}}^{**}$ $(18.9)$ & ${8.4 \times 10 ^ {2}}^{**}$ $(14.2)$ \\
		Neighbours	&  ${9.5 \times 10 ^ {-4}}^{**}$ $(2.1 \times 10 ^ {-5})$ & ${9.5 \times 10 ^ {-4}}^{**}$ $(1.7 \times 10 ^ {-5})$& ${9.2 \times 10 ^ {-4}}^{**}$ $(1.5 \times 10 ^ {-5})$\\
		Infected Communities &  ${0.04}^{**}$ $(0.02)$ & ${0.05}^{**}$ $(0.01)$ & ${0.06}^{**}$ $(0.01)$ \\
		$H^r$		        &  $-0.85^{**}$ $(0.15)$ & $-0.84^{**}$ $(0.12)$ & $-0.86^{**}$ $(0.10)$ \\
		Intra-community  &  $-0.02$ $(0.17)$ & $-0.11$ $(0.13)$ & $-0.13$ $(0.11)$ \\
		\hline \hline 
	\end{tabular}
	
	\label{table:logitSimLam}
\end{table}

\clearpage

\section*{Acknowledgements}
We thank the Ohio Supercomputer Center for their assistance. W.L. was supported by EPSRC Early Career Fellowship in Digital Economy (Grant No. EP/N006062/1). Z.Z. was supported by the Major Program of National Natural Science Foundation of China (Grant No.\ 11290141), Fundamental Research of Civil Aircraft Grant No.\ MJ-F-2012-04. P.J.M. was supported by the Eunice Kennedy Shriver National Institute of Child Health \& Human Development of the National Institutes of Health under Award Number R01HD075712. The content is solely the responsibility of the authors and does not necessarily represent the official views of any of the agencies supporting this work.
\bibliography{scibib}

\end{document}